# International Coordination and Support for SmallSat-enabled Space Weather Activities

Teresa Nieves-Chinchilla[1], Bhavya Lal[2], Robert Robinson[3], Amir Caspi[4], David R. Jackson[5], Therese Moretto Jørgensen[6], and James Spann[7]

[1] NASA Goddard Space Flight Center, Greenbelt, MD, USA.
[2] IDA Science and Technology Policy Institute, Washington, DC, USA
[3] The Catholic University of America, Washington, DC, USA.
[4] Southwest Research Institute, Boulder, CO, USA.
[5] Met Office, Exeter, UK
[6] University of Bergen, Bergen, Norway.
[7] NASA Headquarters, Heliophysics Division, Washington, DC, USA.

Corresponding author: Teresa Nieves-Chinchilla (teresa.nieves@nasa.gov)

**Key Points:**

- Space weather science and SmallSat technology have matured in parallel, but better international communication and coordination is needed

- International agreements must address orbital debris, spectrum management, export control, launch opportunities, data access, and more

- Challenges described in this commentary point to the need for a permanent international working group to coordinate efforts

**Abstract**

Advances in space weather science and small satellite (SmallSat) technology have proceeded in parallel over the past two decades, but better communication and coordination is needed among the respective worldwide communities contributing to this rapid progress. We identify six areas where improved international coordination is especially desirable, including: (1) orbital debris mitigation; (2) spectrum management; (3) export control regulations; (4) access to timely and low-cost launch opportunities; (5) inclusive data policies; and (6) education. We argue the need for internationally coordinated policies and programs to promote the use of SmallSats for space weather research and forecasting while realizing maximum scientific and technical advances through the integration of these two increasingly important endeavors.

# 1 Introduction

Addressing the science and sociatal challenges related to space weather is a global enterprise, not only because the impacts can be worldwide, but also because the observations required for effective forecasting and specification have international implications (National Science and Technology Council, 2019). This is particularly true as small satellite ("SmallSat") technologies are increasingly being applied to the space weather priorities of many countries. While some international coordination is happening bilaterally between government agencies of individual states, and while such initiatives are important in promoting the utilization of small satellites, there is need for larger, global-scale, multi-lateral coordination. Just as international coordination and cooperation have been adopted for maritime and air transportation systems, similar policies and agreements must now be formulated for space-based observing platforms.

SmallSats are a class of spacecraft with masses typically below 200 kg (with some exceptions), including CubeSats with masses of ~1–10 kg and volumes measured in "units" of ~$10\times10\times10$ cm$^3$ cubes. There is rich literature describing SmallSat capabilities (e.g., National Academies of Sciences, Engineering and Medicine [NASEM] 2016; Lal et al., 2017; Millan et al., 2019). A recent report noted, "these lower-cost satellites' expendability, faster refresh, and simultaneous deployment in large numbers—to enable lower-cost spatially or temporally distributed data collection—enables greater risk-taking, experimentation, and creation of new applications not feasible with larger satellites" (Lal et al., 2017). As a result, SmallSats have made forays in almost every area of space, including science and exploration. Multiple prior missions exemplify the feasibility of using SmallSats for high-quality space weather-related research (Spence et al., 2020), while new missions promise to further expand these capabilities (Caspi et al., 2020).

Per NASEM (2016), one of the most promising potentials for CubeSats in science is that they enable launching "low-cost constellations and swarms comprising hundreds or even thousands of data collection platforms," thereby introducing "entirely new architectures and ways to conceptualize space science." A COSPAR-sponsored international study had similar findings (Millan et al., 2019). Because of the vast domain over which space weather occurs, spanning from the Sun to the Earth's surface and beyond, extended SmallSat constellations are particularly desirable for space weather research and monitoring.

Accurate terrestrial weather prediction achieved great advances through deployment of a comprehensive observation network, and the same strategy is required to realize significant advances in space weather prediction capabilities.

Here, we identify several significant challenges posed by the blossoming deployment of SmallSat constellations that require international coordination and policy responses. We briefly summarize each challenge and conclude with recommendations on how the international space weather community may begin to address them. Whereas we focus here on issues requiring international coordination, other papers in this special issue highlight technological challenges and opportunities in applying SmallSats to achieve space weather goals (e.g., Caspi et al., 2020; Verkhoglyadova et al., 2020).

## 2 Challenges in SmallSat Development, Launch, and Flight Operations

*2.1 Orbital Debris*

SmallSat use is growing quickly: in 2019 alone, almost 250 satellites with mass under 200 kg were launched, more than 5 times the number of SmallSats launched during 2012 (Bryce Space, 2020). More than 1,100 CubeSats were launched during 2012–2019. While not all remain in orbit, they add to the existing 900,000 pieces of orbital debris larger than a marble, with 34,000 larger than a softball (i.e., about the size of a 1U CubeSat), tracked by government agencies (ESA, 2020). As the number of SmallSats in orbit increases, the concern is not just the growing number of satellites occupying orbital space, but also that (depending on their altitude) they will likely stay aloft as debris beyond their useful life, presenting a collision hazard for human spaceflight and for other satellites and robotic missions (Berger et al., 2020). The growing number of small objects in orbit can also threaten the integrity of ground-based astronomical observations (Witze, 2019).

Despite the scientific, economic, and other benefits of SmallSat launch and use, these threats from the growing number of SmallSat constellations represent a global problem. Going forward, assuming that not all SmallSats can maneuver, international policies and restrictions will be imperative to ensure that all SmallSats: (1) can be tracked, either actively or passively; (2) cause no radio frequency interference (see next section); and (3) abide by stricter guidelines to de-orbit after they stop functioning**.** For example, the United Nations Committee on the Peaceful Uses of Outer Space (UNCOPUOS, 2019; see Annex II, Guideline B.8) has established policies for tracking and de-orbiting of small satellites. However, international guidelines have not been codified into law, and compliance rates remain low.

*2.2 SmallSat Communications*

Communications are a particular bottleneck for space weather operations, whether from single SmallSats or from constellations (Hapgood, 2008). As discussed below, frequency licensing for radio communications is a complicated and lengthy process even for highly-experienced mission teams. Frequency licensing and coordination is necessarily international, since many SmallSats transmit telemetry nearly continuously while crossing over dozens of countries each orbit. Additionally, spectrum licensing agencies may impose bandwidth restrictions for certain radio-frequency (RF) bands depending on the type of mission (e.g., Earth-imaging versus celestial imaging or *in situ* measurements) regardless of the actual data volume that mission may require. In low-Earth orbit (LEO), visibility of ground stations will limit downlink capacity and hence data "timeliness" (latency). Adding ground passes to boost downlink capacity or reduce latency may not be feasible or affordable, and other solutions must be investigated.

While spectrum-related issues are qualitatively similar for SmallSats as for larger spacecraft, the speed with which SmallSats, especially CubeSats, can be developed and launched is outpacing current coordination processes for spectrum allocation and management. Procedures for receiving permission for spectrum use are long, complicated, and in many countries, spread across multiple agencies. Many researchers deploying science-related CubeSats are unfamiliar with these rules and regulations, and sometimes discover them late in the development process, risking denial of a license or, worse, launching without a license, as was the case with the American startup Swarm Technologies (Harris, 2018). CubeSat developers have historically favored lower frequencies (e.g., UHF or S-band), where equipment is less expensive and more readily available, but these are also the most congested parts of the radio spectrum. The growing use of CubeSats and the accompanying explosion in data volume increases the need for higher bandwidth, which has its own set of costs and challenges (as discussed earlier). Regulatory authorities also prefer to know details of satellite orbits when spectrum filings are made, but these parameters may be uncertain until late in the process, particularly for SmallSats launched as secondary payloads where the primary may not be known until only ~12 months before launch. This challenge is exacerbated for international and joint projects where spectrum allocations of multiple countries may need to be aligned.

In the next decade, if all proposed constellations are launched (an unlikely scenario but worth considering), up to 20,000 satellites could be launched into LEO, most of them under 500 kg (Maclay et al., 2019). This rapid proliferation of SmallSats places increasing pressure on coordination in UHF, S, and X bands as well as other space-allocated bands, since many commercial operators use the same spectral bands as university or federal government agencies. As more satellites are launched, the competition for bandwidth will intensify, not just among satellites in LEO but also in GEO, and in some situations between GEO and LEO satellites. As RF interference becomes more of a problem, enforcement of national and international regulations to prevent interferences will increase, challenging the science community to continually stay apprised of changes. The International Telecommunications Union (ITU) has implemented procedures specifically aimed at regulating bandwidth for SmallSat communication and telemetry (von der Ohe, 2020).

Optical laser communication is an emerging technology with over 100,000 times more frequency bands than traditional RF, operates at lower power levels, can achieve much higher data rates, and can potentially be lighter and smaller (Klumpar et al., 2020). The main problem for laser communication is cloud cover that can block transmissions, but it may also be a valuable capability for communications between spacecraft.

*2.3 Export Controls*

Recent advances in miniaturization of critical spacecraft systems enable SmallSats as viable and cost-effective platforms for space weather research. These include high-precision attitude determination and control systems (ADCSs) for accurate three-axis stabilized pointing; high-powered and resilient processors for on-board data processing and sophisticated command handling; increased mission lifetime; and high-speed, high-bandwidth communications using S- and X-band radio frequencies. Improved efficiency of space-rated multi-junction photovoltaic solar cells and innovations in miniature panel deployment and articulation enable high power generation from a relatively small footprint.

New technologies are under development to enable large SmallSat constellations, particularly ones requiring interaction between spacecraft, and to improve both data speeds and latency. Some of these are discussed in the companion paper by Klumpar et al. (2020). For example, miniaturized propulsion technology provides station-keeping capabilities for SmallSats, whether to combat orbital decay to improve mission lifetime or to enable large constellations whose constituent spacecraft must maintain a known and constant configuration/separation. Many options are becoming commercially available, including cold gas thrusters and ion propulsion, but have not yet been commonly adopted. These innovative technologies represent intellectual property subject to control by individual nations, potentially impeding the international partnering that is the hallmark of many SmallSat missions.

While export control regulations typically exclude general scientific, mathematical, or engineering principles in the public domain (e.g., basic and applied research), they are often hard to interpret by scientists. In some countries, concepts such as "deemed exports" – items or information provided to a foreign individual – are often difficult to understand and follow, and responsibility for complying with these laws often resides with researchers and students not trained in such matters. There is ongoing debate between government and academia regulated by export controls regarding the extent to which these restrictions harm scientific activity. Institutions of higher education in the United States argue that overly hawkish export control regulations inhibit the best international students from studying in the U.S. and prevent cooperation on international projects. Over time, export control-related laws and regulations have become more complicated and more aggressively enforced by government agencies. In the U.S., where enforcement information is publicly available, university personnel have been prosecuted for breaches. Despite recent changes to U.S. policy that now place many export controls for "pure research" missions under the Department of Commerce rather than the State Department, this remains a driving concern. Harmonizing international collaborations while ensuring export control compliance of their research has become a precarious balancing act for scientists.

*2.4 SmallSat Launch Opportunities*

SmallSat developers globally are increasingly looking for low-cost launch opportunities wherever they can find them (e.g., Frick & Niederstrasser, 2018). Nearly 150 small launchers either already exist or are being developed for launching SmallSats (Niederstrasser & Madry, 2020). Historically, however, it has been more economical for SmallSats to launch as rideshares on larger rockets. In 2017, India's PSLV rocket launched over 100 satellites developed by the U.S., the Netherlands, Israel, Kazakhstan, and Switzerland. International cooperation and coordination will facilitate future efforts and expand available options for SmallSat deployment and operation. For instance, the Access to Space for All initiative of the United Nations Office for Outer Space Affairs (UNOOSA) connects established and emerging space actors to ensure the benefits of space are available to non-spacefaring and emerging space-faring nations.

*2.5 Data Policies*

An open data policy is desirable to maximize the scientific and operational impact of new SmallSat data (e.g., on space weather forecasts). For operational applications, it may be worthwhile to follow the World Meteorological Organization (WMO) and other bodies (e.g., European Organisation for exploitation of Meteorological Satellites [EUMETSAT]) in defining a list of "essential" data and products that would be made available world-wide on a free and unrestricted basis. UNCOPUOS has also established policies for sharing of space weather data (UNCOPUOS, 2019, Annex II). Standardization of space weather data products will facilitate data exchange and ease of use. However, data sharing should allow researchers to retain preferential access to more innovative observations and ample opportunity to exploit those data via peer-reviewed publications. As with most missions, there should be a period immediately after launch reserved for calibration when data need not be shared and, clearly, the instrument developers should own Intellectual Property Rights for the data they create. To provide further incentive for this open data policy, funding agencies should require a meaningful "pathway to impact" for SmallSat data and can develop frameworks to facilitate these pathways by, for example, funding near-real-time downlinks for operational use, or supporting missions that are demonstrators for future operational missions.

*2.6 SmallSat Educational Efforts*

The educational aspects of SmallSats (particularly CubeSats) are an intrinsic part of their heritage. CubeSats, with their concomitant philosophy of standardization and containerization, were created in 1999 at California Polytechnic State University (CalPoly) and the Space Systems Development Lab at Stanford University to facilitate access to space for university students at low cost (see https://www.cubesat.org/). The U.S. National Science Foundation (through the "CubeSat-based Science Missions for Geospace and Atmospheric Research" program) advocated for this concept and served as a trigger to motivate the flow of new ideas from academia to the scientific community (Moretto & Robinson, 2008).

In the last few years, the scientific community has been actively working to prepare our technologically advanced societies to reduce vulnerabilities to space weather hazards. Among other roles, the community advances knowledge of the fundamental nature of space weather, contributes to developing a reliable space weather prediction and forecast system, and evaluates space weather effects on human and technological assets. Long-term planning requires a multidisciplinary approach, with efforts to promote the flow of knowledge from the scientific community towards academia. SmallSat capabilities and legacy are key pieces in both requirements.

Currently, there is a significant lack of space weather programs within U.S. colleges and universities and lack of a critical mass of students and faculty within relevant departments. Space weather is inherently interdisciplinary, but researchers interested in space weather are typically trained in academic departments that lack the breadth and depth that the field demands. The cost to create a space weather program in universities is very high and the number of potential students is likely low. This scenario has created challenges in overcoming the scientific and technical complexities of space weather research, and in the practical problem of sustaining support for space weather-related infrastructure and human resources. Although significant effort is being made by the scientific community (e.g., the Community Coordinated Modeling Center), ensuring the transfer of knowledge requires creating an international collaboration between academia, industry, government agencies, and international organizations to promote cooperative academic programs that minimize the cost and maximize the number of end-users.

## 3 Conclusions and Recommendations

As the market and demand for SmallSats grow and their use becomes more commonplace, these platforms become more capable for implementing space weather research and operations in regions of parameter space that can be entirely inaccessible by traditional larger missions, and at lower cost. Much of this advancement has, to date, been driven by the commercial market, but is typically sponsored by grants or contracts from government agencies. These agencies should continue to recognize the need for innovation and sponsorship of opportunities in this growing market. However, sharing technological advancements between different nations is complicated by the challenges outlined above. These will need to be overcome to enable increased international collaborations in SmallSat-driven space weather research and operations.

The challenges described in this commentary point to the need for a permanent international working group to coordinate efforts, produce and maintain a list of best practices for SmallSat developers, and recommend regulations that will guide future SmallSat operations. Schrijver et al. (2015) developed a space weather roadmap including recommendations for future Smallsat-based space weather observations. Following from this, COSPAR commissioned an international study group to construct a SmallSat Roadmap (Millan et al., 2019). Recommendations were aimed at scientists, industry, space agencies, policy makers, and COSPAR, with an underpinning aim of increasing exploitation of SmallSats, and increasing flexibility to ensure their application to space weather. They suggest COSPAR could facilitate international teams to come together like in the QB50 project (e.g., Gill et al., 2013) to meet large-scale science goals via SmallSat constellation missions. This is not inconsistent with our recommendation for an international working group, but misses our further aim of enabling SmallSats to ultimately benefit providers of operational space weather services and the end users of such services.

As discussed in the companion paper by Verkhoglyadova et al. (2020), the WMO produces requirements for space weather observations with an emphasis on near-real-time operations. The existing observational network is regularly assessed against these requirements to identify gaps in provision and to advocate for developments in the network. These efforts need to be better coordinated with other groups of observation providers, such as the Coordination Group for Meteorological Satellites (CGMS), and the SmallSat community, especially since the gaps between provision and requirements are currently often very large.

Another goal of this improved coordination is to better publicize operational space weather observational needs. A longer-term aim is to strike a balance between the WMO requirements – designed to meet the needs of the users of operational space weather services rather than necessarily being linked to upcoming observational developments – and research into new observational methods being carried out by the SmallSat community and other researchers. This balance is essential to ensure a strong connection between research and operations, to enable a pathway for continual research-to-operations developments, and to minimize the risk of lack of engagement (e.g., via the researchers dismissing the WMO requirements as being too challenging). An effective way of achieving this connection is for our proposed working group to organize a research-to-operations observations workshop, jointly sponsored by stakeholder agencies worldwide.

**Acknowledgements**

A.C. was partially supported by NASA grants NNX14AH54G, NNX15AQ68G, NNX17AI71G, and 80NSSC19K0287. The authors thank the organizers of the 1$^{st}$ International Workshop on Small Satellites for Space Weather Research & Forecasting (SSWRF). No new data were used in preparing this manuscript.